\nonstopmode
\documentclass[fleqn,usenatbib]{mnras}
\usepackage{newtxtext,newtxmath}
\usepackage[T1]{fontenc}

\DeclareRobustCommand{\VAN}[3]{#2}
\let\VANthebibliography\thebibliography
\def\thebibliography{\DeclareRobustCommand{\VAN}[3]{##3}\VANthebibliography}

\usepackage[dvipsnames]{xcolor}
\usepackage{graphicx}
\usepackage{amsmath}
\usepackage{hyperref}

\hypersetup{colorlinks, citecolor = [rgb]{0,0.5,0} }

\title[CF4 Velocity Correlation]{The peculiar velocity correlation function of the Cosmicflows-4 catalog}

\author[Y.~Wang et al.]{
Yuyu Wang$^{1,2}$\thanks{E-mail: yuyuwang@ohio.edu},
Hume A. Feldman$^{3}$,
Richard Watkins$^{4}$,
and Xiaohu Yang$^{1,5}$
\thanks{E-mail: xyang@sjtu.edu.cn},
\\
$^{1}$Department of Astronomy, School of Physics and Astronomy, and Shanghai Key Laboratory for Particle Physics and Cosmology, Shanghai Jiao Tong \\
University, Shanghai, 200240, China.\\
$^{2}$Department of Physics \& Astronomy, Ohio University, Athens, OH, 45701, USA.\\
$^{3}$Department of Physics \& Astronomy, University of Kansas, Lawrence, KS 66045, USA.\\
$^{4}$Department of Physics, Willamette University, Salem, OR 97301, USA.\\
$^{5}$State Key Laboratory of Dark Matter Physics, Tsung-Dao Lee Institute, Shanghai Jiao Tong University, Shanghai 201210, China.
}

\date{Accepted XXX. Received YYY; in original form ZZZ}
\pubyear{2025}

\begin{document}
\label{firstpage}
\pagerange{\pageref{firstpage}--\pageref{lastpage}}
\maketitle

\begin{abstract}
We present an analysis of the parallel peculiar velocity correlation function using data from the Cosmicflows-4 (CF4) survey. CF4 significantly extends the depth of the peculiar velocity measurements, mitigating the impact of observers on the cosmic variance. We examine the distribution of cosmic variance using different velocity correlation estimators. The combination of the large peculiar velocity uncertainties and the anisotropy distribution of the CF4 data across the northern and southern hemispheres results in substantial statistical uncertainties in the velocity correlation function. To address this, we test different weighing schemes in the velocity correlation function and implement a more accurate peculiar velocity estimator that reduces velocity uncertainties, consequently decreasing the statistical uncertainty. Using the CF4 group dataset, we derive a growth rate of $f\sigma_8=0.384^{+0.116}_{-0.194}$ and a local growth rate of $f\sigma_8=0.569^{+0.054}_{-0.06}$ through a Markov Chain Monte Carlo method.
\end{abstract}

\begin{keywords}
large-scale structure of Universe -- cosmological parameters -- galaxies: peculiar -- methods: statistical
\end{keywords}

\section{Introduction}\label{sec:intro}

Studies of peculiar velocities provide an independent perspective on the large-scale structure of the universe. Unlike the galaxy distribution, the peculiar velocity is an unbiased tracer of the large-scale structure \citep[e.g.][]{WatFelHud2009, FelWatHud2010, NusBraDav2011, MacFelFer2011, MacFelFer2012, Nusser2014, SprMagColMou2014,  GraCouLav2019, Howlett2019, FerBun2022, DupCou2023, PriLecLav2023, ZhaPerDin2024, ShiZhaMao2024, ValLibPom2024, CouMouHol2025, QinBlaHow2025, LaiHowAgu2025}. However, peculiar velocity studies are constrained by the survey depth and precision, limiting the research to nearby redshift regions ($z<0.1$). In contrast, studies employing mass distributions have extended to much deeper redshifts \cite[e.g.][]{KerSzaSza2000, VarBauHam2013, WanBruDol2013, CacVosMor2013, ManSloBal2013, KeiKurLin2019, XuLiZha2023, LiuZhaLi2023}.

The primary challenge in measuring peculiar velocities lies in obtaining accurate distance information \citep[e.g.][]{deJBelChi2012, CF3, TulKouCou2023, MouJarCou2024, SaiHowDav2024}, which directly impacts velocity precision. Conventional methods like the Tully-Fisher relation \citep[TF,][]{TullyFisher1977} and the Fundamental Plane \citep[FP,][]{DjoDav1987,DreLynBurDav1987} are commonly used for distant galaxy detections, which suffer from fractional errors around 20\%. This large distance uncertainty translates into more significant uncertainties in peculiar velocity measurements. In addition, the distribution of distance errors is not Gaussian due to the logarithm relation between distances and distance modulus. This leads to non-Gaussian peculiar velocity uncertainties, which can be mitigated by using modified distance or velocity estimators \citep[e.g.][]{NusDav1995, WatFel2015a, Sorce2015, HofNusVal2021, WatAllBra2023}. In this paper, we implement the distance estimator introduced by \citet{WatAllBra2023}

For more distant galaxies, distance measurement becomes even more challenging. Consequently, peculiar velocity predictions in these higher redshift regions rely on distance-independent methods, such as velocity reconstruction \cite[e.g.][]{DavNusMas2011, WanMoYan2012, Lavaux2016, KesNus2017, SorHofGot2017, YuZhu2019, ZhuWhiFer2020, BorLavHud2022, ValLibHof2023, BayModFer2023, TurBla2023, QinParHon2023, HofValLib2023, WanYan2024, BlaTur2024, BroGel2024, HofValLib2024, ShiWanYan2025} and the kinetic Sunyaev-Zeldovich effect \citep[e.g.][]{SunZel1980, RepLah1991, DolHanRon2005, BhaKos2008, KasAtrKoc2008, KasAtrKoc2009,  HanAddAub2012, PlanckAde2013, DolKomSun2015, SayZemGle2015, PlanckAgh2017, SoeSarGia2017, Hurier2017, KirSav2018, PlanckAgh2019, WanRamSal2021, GonBea2024, LagMadSmi2024, McCBattBea2024, DuShiWan2025}.

To overcome the limitations of large uncertainties in individual peculiar velocity measurements, several ensemble statistics of peculiar velocities have been introduced to explore the universe, such as Bulk Flow \citep[e.g.][]{Kaiser1988, FKP1994, WatFel1995, WatFelHud2009, FelWatHud2010, AgaFelWat2012, Nusser2014, KumWanFelWat2015, ScrDavBla2015, SeiPar2016, HofNusCorTul2016, Nusser2016, PeeWakFel2018, QinHowStaHon2019, Qin2021, WhiHowDav2023, WatAllBra2023, AviOliDia2023, LopBerFra2024} that is generally defined as the average of peculiar velocities in a volume, Pairwise Velocity \citep[e.g.][]{FerJusFel1999, JusFerFelJaf2000, FelJusFer2003, ZhaFelJus2008, HanAddAub2012, Hellwing2014, PlanckXXXVII2015, MaLiHe2015, CheZhaYan2022, XiaZhe2023, LiZheChe2024, CheZha2024}, which indicates the mean difference of peculiar velocity pairs of different separations. 

In this paper, we focus on the Velocity Correlation Functions \citep[e.g.][]{JafKai1995, ZarZehDekHof1997, JusFerFelJaf2000, BorCosZeh2000,  AbaErd2009, NusDav2011, OkuSelVla2014, HowStaBla2017, HelNusFei2017, WanRooFel2018, DupCouKub2019, WanPeeFel2021, TurBlaRug2021, TurBlaRug2023, LyaBlaTur2024}, which describe the velocity correlation of galaxy pairs. The velocity correlation function has been widely used in cosmological studies. For example, \citet{DupCouKub2019} obtained $f\sigma_8=0.43\pm0.03$ using the Cosmicflows-3 \citep[CF3,][]{CF3} dataset. \citet{WanPeeFel2021} constrained $\Omega_m$ and $\sigma_8$ using the CF3 data. In addition, some studies \citep[e.g.][]{TurBlaRug2021, LyaBlaTur2024, TurBlaQin2025} improved constraints on $f\sigma_8$ through the joint analysis of galaxy-galaxy auto-correlation, galaxy-velocity cross correlation and velocity auto-correlation. More recently, \citet{TonAppPar2024} constrained the growth rate to $f\sigma_8=0.361^{+0.02}_{-0.027}$ using simulation data.

Previous work by \cite{WanPeeFel2021} introduced a volume-weighted velocity correlation function estimator to mitigate the correlation bias caused by the non-Copernican observer, which is discussed in \cite{HelNusFei2017}. Although the correlation bias and cosmic variance are improved with the new estimator, the volume-weighted velocity correlation function suffers from significant statistical uncertainty. In this paper, we present a novel correlation estimator using the largest peculiar velocity catalog, Cosmicflows-4 \citep[CF4,][]{TulKouCou2023}. Compared to its predecessor, Cosmicflow-3 \citep[CF3,][]{CF3}, CF4 significantly expands the survey depth from $z\sim 0.05$ to $z\sim 1$ and nearly triples the galaxy sample size (from about 18,000 to 55,877). We expect both correlation bias and statistical uncertainty to be mitigated with the large sample size and extensive sky coverage of CF4. However, CF4 exhibits an anisotropic galaxy distribution in the northern and southern hemispheres. To address the potential bias caused by this imbalance, we introduce a kernel density weighting scheme to the velocity correlation function analysis in this paper.

The paper is organized as follows. Section~\ref{sec:method} details the derivation of the estimators for the velocity correlation function. In Section~\ref{sec:data}, we introduce the observation and simulation data implemented to calculate the velocity correlation functions. In Section~\ref{sec:result}, we discuss the result of the velocity correlation functions. Section~\ref{sec:constraints} presents the cosmological parameters constrained by the velocity correlation function. We conclude in Section~\ref{sec:conclusion}.

\section{The velocity correlation function}\label{sec:method}

In velocity correlation studies, the Gorski correlation functions $\psi_1$ and $\psi_2$ \citep{Gorski1988, GorDavStr1989} are the most widely used. However, it is important to note that the cosmic variance of $\psi_1$ exhibits a non-Gaussian distribution \citep{WanRooFel2018}, while $\psi_2$ lacks the desired stability when applied to data sets with finite sizes \citep{GorDavStr1989}. In addition, \citet{GorDavStr1989} introduced the parallel correlation function ($\Psi_\parallel$) and the perpendicular correlation function ($\Psi_\perp$), which can be expressed in terms of $\psi_1$ and $\psi_2$. \cite{WanPeeFel2021} introduced new estimators for $\Psi_\parallel$ and $\Psi_\perp$ that can be calculated directly from line of sight peculiar velocities. $\Psi_\parallel$ displays a Gaussian distributed cosmic variance, effectively mitigating the non-Gaussian cosmic variance issue in the Gorski's correlation functions. In contrast, $\Psi_\perp$ is less stable, manifesting non-Gaussian cosmic variance, and is more sensitive to different types of observers, as discussed in section~\ref{sec:result_CV}. Additionally, the parallel and perpendicular correlation functions represent physical correlations that can be calculated directly from three-dimensional peculiar velocities, whereas the Gorski correlation functions are geometry-modified correlations that may be influenced by the survey distribution. In this paper, we employ the parallel ($\Psi_\parallel$) and perpendicular ($\Psi_\perp$) correlation functions to conduct an in-depth analysis of CF4 data.

The estimators of $\Psi_\parallel$ and $\Psi_\perp$ are derived using the least squares method following \cite{Gorski1988} and \cite{Kaiser1989}, with details provided in \cite{WanPeeFel2021}.
\begin{equation}
\hat{\Psi}_{\parallel} (r)= \frac{\sum wg^2 \sum wf u_1u_2- \sum wfg \sum wg u_1u_2}{\sum wf^2 \sum wg^2 - \left( \sum wfg \right)^2 } ,
\label{eq:psipar}
\end{equation}
and
\begin{equation}
\hat{\Psi}_{\perp}(r) = \frac{\sum wf^2 \sum wg u_1u_2 - \sum wfg \sum wf u_1u_2}{\sum wf^2 \sum w g^2 - \left( \sum wfg \right)^2 },
\label{eq:psiperp}
\end{equation} 
where the sums are over galaxy pairs whose separations lie in a bin centered on redshift-space distance $r$, $w$ is a weight assigned to each galaxy pair, which is set to be 1 in all subsequent calculations, expect in Section~\ref{sec:weight}. $f=\cos(\theta_1)\cos(\theta_2)$, $g=\sin(\theta_1)\sin(\theta_2)$, the quantities $u_1$ and $u_2$ are the line of sight peculiar velocities in a given pair. $\theta_1$ and $\theta_2$ represent the angles formed between the  separation vector $\textbf{r}$ of a galaxy pair and the respective position vectors $\textbf{r}_1$ and $\textbf{r}_2$ of the two galaxies. These angles can be expressed by $\cos\left(\theta_1\right)=\left(\widehat{\textbf{r}}\cdot\widehat{\textbf{r}}_1\right)$ and $\cos\left(\theta_2\right)=\left(\widehat{\textbf{r}}\cdot\widehat{\textbf{r}}_2\right)$, where $\textbf{r} = \textbf{r}_1 - \textbf{r}_2$. Due to the large uncertainty in distance measurements, the separation is calculated by redshift distance in this paper, which is given by $cz/100$ in units of h$^{-1}$ Mpc. 

The weights $w$ in the estimators of $\Psi_\parallel$ and $\Psi_\perp$ are assigned differently depending on the purpose. \cite{Kaiser1989} set the weight to 1, while \cite{GroJusOst1989} chose weights designed to mitigate the effects of measurement errors. In \cite{WanPeeFel2021}, we introduced a volume weighting scheme to address the bias caused by different observers \citep{HelNusFei2017}, defined as $w=(r_1r_2)^p$, where p takes values of 0.5, 1, and 2 (with p=0 indicating uniform weights). It is notable that the values p = 0.5, 1, and 2 were adopted as representative test, rather than as physically optimized values. Following \citep{HelNusFei2017} that the cosmic variance of velocity correlation functions can depend on the choice of observers, the volume-weighting scheme was designed to test weather the observer dependence seen in shallow surveys (e.g., CF3) is driven by the strong radial decline in galaxy number density. This weighting scheme up-weights more distant galaxies to reduce the artificial dominance of nearby regions, therefore mitigating this observer-related bias in cosmic variance. The observer challenges posed by the unbalanced galaxy distribution can be further improved with more extensive velocity surveys that have larger sample sizes and greater depth. Compared to CF3, CF4 contains more peculiar velocities and spans a larger redshift region. Therefore, we expect the velocity correlation functions of the CF4 to be less sensitive to the observers and adopt p=0 for CF4 analyses, which is discussed in detail in section~\ref{sec:result}.

Furthermore, \cite{TurBlaRug2021} introduced the random catalog, a common practice in studies of galaxy two-point correlation functions, into the calculation of the Gorski's velocity correlation function and the density-velocity cross-correlation function. The inclusion of random catalogs significantly reduced the uncertainty in the density-velocity cross-correlation function; however, its impact on the velocity correlation function was negligible. Unlike the galaxy distribution, the peculiar velocity serves as an unbiased tracer of the mass distribution, which makes the velocity correlation function minimally affected by the inclusion of random catalogs.

In linear theory, $\Psi_\parallel$ or $\Psi_\perp$ are defined by
\begin{equation}
\Psi_{\|}(r)=\frac{\left(f H_0\right)^2}{2 \pi^2} \int P(k)\left[j_0(k r)-2 \frac{j_1(k r)}{k r}\right] d k,
\end{equation}
\begin{equation}
\Psi_{\perp}(r)=\frac{\left(f H_0\right)^2}{2 \pi^2} \int P(k) \frac{j_1(k r)}{k r} d k,
\end{equation}
where $f=\Omega^{0.55}_m$ \citep{Linder2005}, $H_0$ is the Hubble constant, $j_n(x)$ are the spherical Bessel functions, and the matter power spectrum $P(k)$ is modeled using the formula of \citet{EisHu1998}. By fitting the velocity correlations obtained through linear theory with those from the estimators, one can constrain the cosmological parameters.

\section{Data}\label{sec:data}

\subsection{Observational data from Cosmicflows-4}\label{sec:sub_data_cf4}

The Cosmicflows-4 \citep[CF4,][]{TulKouCou2023} is the latest version of the Cosmicflows program, which compiles galaxy distances from different observation surveys for peculiar velocity analysis. The CF4 catalog contains 55,877 galaxies, including galaxies from Cosmicflows-3 \citep[CF3,][]{CF3}, which is a compilation of Type Ia supernovae \citep{TonSchBar}, Tully-Fisher (TF) spiral galaxy clusters \citep{GioHaySal1998, DalGioHay1999}, Streaming Motions of Abell Clusters (SMAC) from FP survey \citep{HudSmiLuc1999, HudSmiLuc2004}, FP early-type far galaxy clusters \citep{ColSagBur2001}, TF clusters \citep{Willick1999}, the SFI++ catalog \citep{MasSprHay2006, SprMasHay2007, SprMasHay2009}, the SFI++ group catalog \citep{SprMasHay2009}, an early-type nearby galaxy survey \citep{daCBerAlo2000, BerAlodaC2002, WegBerWil2003}, a surface brightness fluctuation survey \citep{TonDreBla2001}, Spitzer Space Telescope \citep{SorTulCou2014}, and 6-degree Field Galaxy Survey \citep{SprMagCol2014}, together with additions from Arecibo Legacy Fast ALFA survey \citep{HayGioMar2011, HayGioKen2018} and Sloan Digital Sky Survey \citep[SDSS,][]{SDSS2000, HowSaiLuc2022}. Furthermore, the inclusion of data from the SDSS provides FP distances in the northern hemisphere and extends the CF4 to the redshift of $z\sim 0.1$.

These 55,877 galaxies are clustered into 38,065 groups, forming the CF4 group dataset. Most of these groups are defined based on scaling relations between a proxy for the virial radius, velocity dispersion, and mass \citep{Tully2015, KouTul2017}. Notably, the CF4 group dataset provides measurements with uncertainties 20\%-25\% lower than those observed in individual galaxy cases. In this paper, we utilize the CF4 group dataset to conduct our analysis of the velocity correlation function.

\begin{figure}
\centering
\includegraphics[width=8.5cm]{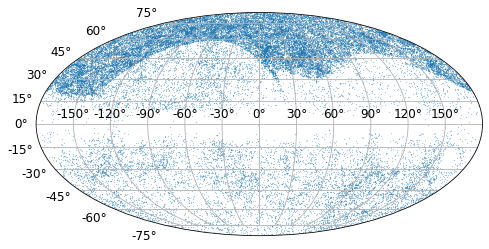}
\caption{\label{fig:CF4-group} The angular distribution of CF4 groups in galactic coordinates.}
\end{figure}

\begin{figure}
\centering
\includegraphics[width=8.5cm]{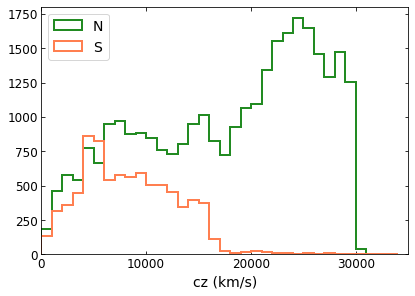}
\caption{\label{fig:CF4-NS-radial} The redshift distribution of CF4 groups in the northern and southern hemispheres. The green histogram bars indicate the redshift distribution in the northern hemisphere, while the orange bars represent the distribution in the southern hemisphere.}
\end{figure}

The addition of CF4 is mainly concentrated in the northern sky, resulting in an unbalanced distribution, as depicted in Fig.~\ref{fig:CF4-group}. Compared to the southern hemisphere, there are more groups distributed in the northern hemisphere, making the CF4 anisotropic. In addition, the survey depth in the northern sky is significantly deeper due to the SDSS data, reaching approximately $z\sim 0.1$, in contrast to the southern sky depth of $z\sim 0.05$, as illustrated in Fig.~\ref{fig:CF4-NS-radial}. The figure exhibits a disparity in the number of groups between the northern and southern hemispheres, with approximately 29,900 groups in the northern sky compared to about 8,000 in the southern sky.

\cite{WatAllBra2023} introduces an unbiased distance estimator for the CF4 data to correct the non-Gaussian distance uncertainty caused by the logarithmic relation between the distance modulus and the distance, as expressed by
\begin{equation}\label{eq:corrected_distance}
d_c = 10^{\frac{\mu}{5}-5}exp(-(\kappa\sigma_\mu)^2/2),
\end{equation}
where $d_c$ indicates the unbiased distance, $\mu$ is the distance modulus and $\kappa=ln(10)/5$. The corresponding distance uncertainty is given by
\begin{equation}\label{eq:dc_uncertainty}
\sigma_c=\kappa\sigma_\mu10^{\frac{\mu}{5}-5},
\end{equation}
where $\sigma_\mu$ represents the uncertainty of the distance modulus.

We adopt the peculiar velocity estimator introduced by \citet{DavScr2014} (hereafter referred to as the DS velocity estimator), which incorporates the Doppler Shift and describes the gravitational redshift using a similar framework to that of \citet{SacWol1967}, in order to improve the accuracy of peculiar velocity estimation:
\begin{equation}\label{eq:velocity_hz}
v_{DS} = c(\frac{z-z_l}{1+z_l}),
\end{equation}
where $z$ represents the observed redshift and $z_l$ is the redshift calculated from luminosity distance ($d_c$) by fitting the luminosity distance expression derived from the Friedmann equation. In our analysis, we implement the same form of luminosity distance expression commonly used in many peculiar velocity studies \citep[e.g.][]{MaLiHe2015, HofNusVal2021}:
\begin{equation}\label{eq:dl_to_z}
d_c = (1+z_l) \int ^{z_l}_0 \frac{cdz}{H_0\sqrt{\Omega_m(1+z)^3+\Omega_{\Lambda}}} .
\end{equation}
It is important to note that the $d_c$ values calculated from Eq.~\ref{eq:corrected_distance} and Eq.~\ref{eq:dl_to_z} both represent luminosity distances. Additionally, the distance is used only in the peculiar velocity estimation, correlation separations and mock galaxy selection are performed using redshift-space distance.

In this paper, we implement $d_c$ and $v_{DS}$ for the analysis of velocity correlation functions. According to Eq.~\ref{eq:dc_uncertainty}, the distance uncertainty ($\sigma_c$) of CF4 is approximately 20\%. The uncertainty propagation between $d_c$ and $v_{DS}$ is too complex to drive an analytical equation. Instead, we perturb each of the CF4 distances ($d_c$) by 20\% to generate 100 perturbed CF4 distance datasets. Using these datasets, we then calculate 100 perturbed CF4 peculiar velocity datasets based on Eq.~\ref{eq:velocity_hz} and ~\ref{eq:dl_to_z}. 

\begin{figure}
\centering
\includegraphics[width=8.5cm]{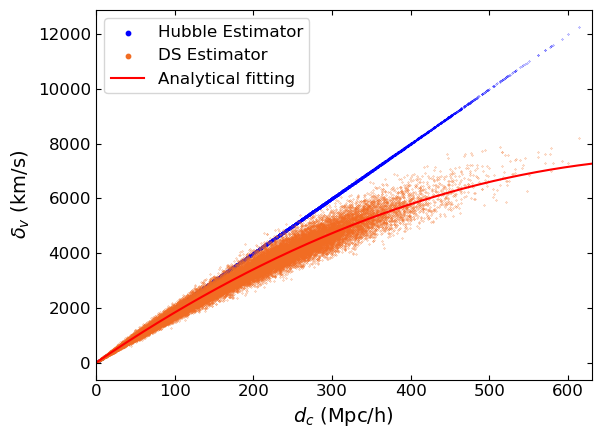}
\caption{\label{fig:CF4-v_error} The uncertainty of CF4 peculiar velocities as a function of luminosity distance. Blue dots represent the velocity uncertainties using the Hubble estimator, orange dots indicate the uncertainties using the DS estimator. The red line displays the analytical fitting results for the DS estimator, respectively.}
\end{figure}

Figure~\ref{fig:CF4-v_error} shows the velocity uncertainty using the DS estimator and the Hubble velocity estimator ($v_H = cz-H_0d_c$). Since the velocity uncertainty is proportional to the distance uncertainty ($20\%d_c$), distant galaxies tend to have higher velocity uncertainties. Compared to the Hubble estimator, the DS estimator yields smaller velocity uncertainties, especially for more distant galaxies. We fit the uncertainty of the DS estimator to an analytical model, resulting in a quadratic function of distance: $\sigma_v = 0.2 H_0d_c (0.98-0.00064 d_c)$. Although this quadratic equation provides a good fit to the CF4 data, it may not be suitable for deeper peculiar velocity surveys. 

Due to the large uncertainties in distance measurements, the uncertainties of peculiar velocities at large distances become extremely large, leading to large statistical uncertainties of velocity correlation functions.

\begin{figure*}
\centering
\includegraphics[width=5.7cm]{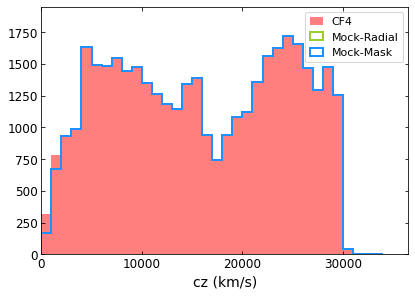}
\includegraphics[width=5.7cm]{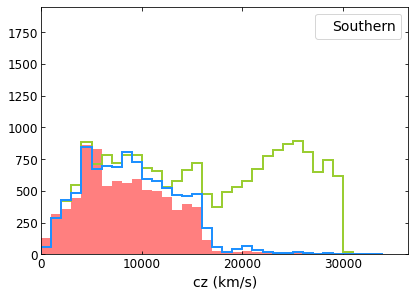}
\includegraphics[width=5.7cm]{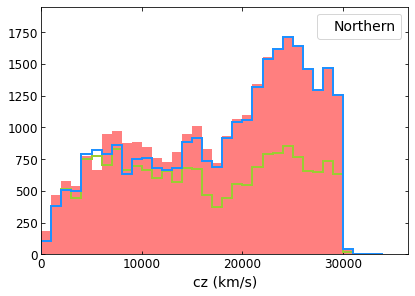}
\caption{\label{fig:CF4-selection} The redshift distribution of the CF4-group observation and simulation catalogs. The red histogram illustrates the redshift distribution of the CF4-group observation catalog, the blue histogram shows the distribution of a Mask mock catalog, while the green histogram displays the distribution of a Radial mock catalog. The left panel indicates the result for the full sky, where the blue and green histograms are overlapped with each other. The middel panel shows the result for the southern hemisphere, while the right panel represents the result for the northern hemisphere.}
\end{figure*}

\subsection{Mock catalogs}\label{sec:sub_data_simul}

Mock catalogs are generated from the halo catalog of the Outer Rim Simulation \citep{HeiFinPop2019}, a dark matter-only simulation with cosmological parameters similar to those of the Wilkinson Microwave Anisotropy Probe 7 \citep[WMAP7,][]{KomSmiDun2011}, as detailed in Table~\ref{T_Mill}. In this paper, we specifically focus on halos within the mass range of $[10^{11}, 10^{14}]$ $h^{-1} M_{\odot}$.

\begin{table}
\caption{Cosmological parameters of the Outer Rim Simulation}
\centering
{
\begin{tabular}{lc}
\hline
Matter density, $\Omega_m$ & 0.2648\\
Cosmological constant density, $\Omega_\Lambda$ & 0.7352 \\
Baryon density, $\Omega_b$ & 0.0448\\
Hubble parameter, $h$ (100 km s$^{-1}$ Mpc$^{-1}$) & 0.71\\
Amplitude of matter density fluctuations, $\sigma_8$ & 0.8\\
Primordial scalar spectral index, $n_s$ & 0.963\\
Box size ($h^{-1}$Mpc) & 3,000\\
Number of particles & $10,240^3$\\
Particle mass, $m_p$ ($10^{9} h^{-1} M_{\odot}$) & 1.85\\
Softening, $f_c$ ($h^{-1}$kpc) & 3\\
\hline
\end{tabular}
}
\label{T_Mill}
\end{table}

The mock catalogs are categorized based on two factors: observers and angular distributions. Concerning observers, we generate mock catalogs with Random (Rand) and Local Group (LG) like observers by the same algorithm described in \cite{WanPeeFel2021}. The Rand-observer mocks are centered on randomly selected halos within the inner (2300 $h^{-1}$Mpc) region of the simulation box, meaning the observer is not positioned in any specific region of the Universe. In contrast, the LG-observer mocks are centered on Milky Way-like halos ($M = [13.5 \pm 6.5] \times 10^{11} h^{-1} M_{\odot}$) with a Virgo-like cluster at a distance comparable to that of the Virgo Cluster from the Milky Way, representing the observer is located in a region similar to our Local Group.

For the angular distribution, the mock catalogs are generated based on two patterns: Radial and Mask. The Radial mocks are created by randomly selecting halos to match the radial redshift distribution of the CF4 catalog, without considering the anisotropy of the CF4 survey. For the Mask pattern, we first generate a CF4-mask by gridding the CF4-group catalog with a grid size of 10 $h^{-1}$Mpc. Then, we select halos within the regions defined by the CF4-mask. The mock catalogs are subsequently generated by randomly choosing the masked halos to match the CF4 radial redshift distribution. 

\begin{figure}
\centering
\includegraphics[width=8.5cm]{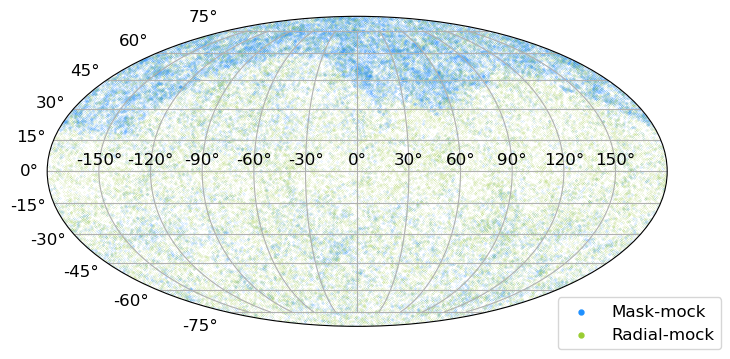}
\caption{\label{fig:CF4-mock} The angular distribution of CF4 mock catalogs. Green dots indicate the halo distribution of a Radial mock catalog, while blue dots represent the distribution of a Mask mock catalog.}
\end{figure}

Fig~\ref{fig:CF4-selection} illustrates the redshift distribution of the CF4-group catalog alongside the Radial and Mask mock catalogs. The left panel displays the redshift distribution for the full sky, the middle panel shows the redshift distribution in the southern sky and the right panel depicts the redshift distribution in the northern sky. In the full-sky figure, both the Radial and Mask mock catalogs match the redshift distribution of the CF4 catalog. However, only the Mask mock catalog agrees with the observation data (CF4) when separating the hemispheres. This indicates that the Mask mock catalog emulates the anisotroy of the CF4 catalog, whereas the Radial mock catalog remains isotropic, as displayed in Fig.~\ref{fig:CF4-mock}.

We generate 1000 catalogs for each type of mocks, finding that the cosmic variance converges adequately with 100 catalogs. Consequently ,we employ 100 mock catalogs in the following analyses and discussions to optimize computational resources. For each type of mock catalog, the average separation between the 100 catalog centers exceeds 1500 $h^{-1}$ Mpc. For the rand-observer mocks, only 0.65\% of center separations fall below 300 $h^{-1}$ Mpc, while this fraction is approximately 0.77\% for the LG-observer mocks. consequently, these 100 catalogs for each mock type can be effectively considered as non-overlapping spheres, ensuring that the analysis of the cosmic variance remains unaffected.  The cosmic variance, denoted as $\delta_c$ in this paper, is estimated by taking the standard deviation of the correlation functions of the 100 mocks. The statistical uncertainty, represented by $\delta_s$, is calculated by perturbing peculiar velocities with uncertainties derived by the quadratic uncertainty equation. The total uncertainty, denoted as $\delta_t$ is defined as the standard deviation of the velocity-perturbed correlations of each mock.

By applying redshift-based selection effect and using redshift distances for correlation separations in both the CF4 observational and simulation data, the impact of Malmquist bias becomes negligible in the analysis of velocity correlation functions for large surveys such as CF4.

\section{Result}\label{sec:result}

\subsection{Cosmic Variance}\label{sec:result_CV}

In \citet{WanRooFel2018}, we demonstrated that the cosmic variance of $\psi_1$ is non-Gaussian when analyzed using CF3 catalog, which describes the variance of velocity correlations across mock catalogs centered at different positions. In this study, we extend the analysis to test the cosmic variance of four velocity correlation estimators ($\Psi_\parallel$, $\Psi_\perp$, $\psi_1$ and $\psi_2$) using the CF4 catalog.

\begin{figure}
\centering
\includegraphics[width=8.5cm]{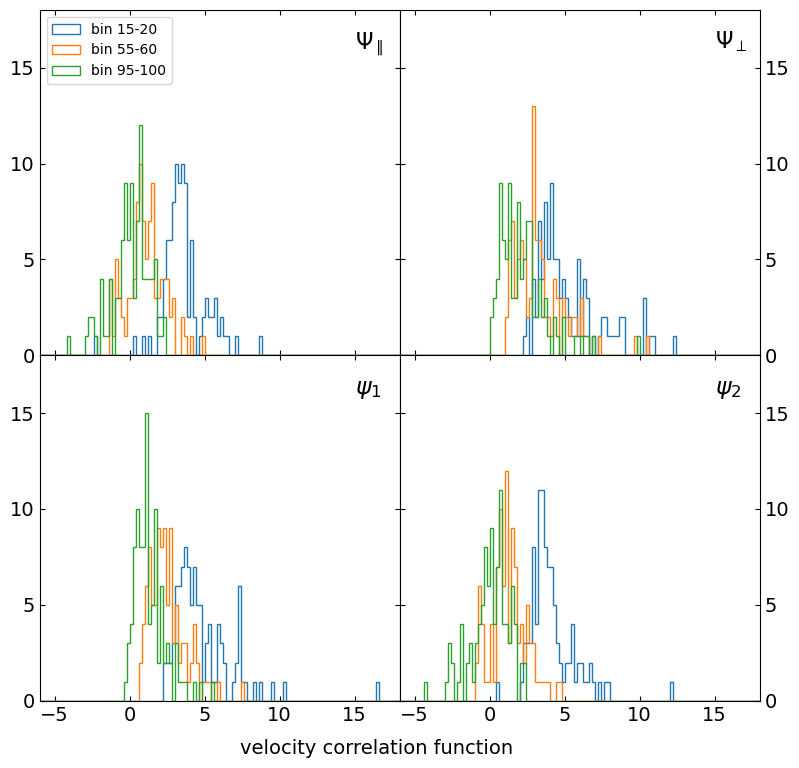}
\caption{\label{fig:CF4-CV_compare} The distribution of velocity correlations in three separation bins of 100 mock catalogs, in units of $(100$ km s$^{-1})^2$. The blue, orange and green histograms indicate cosmic variance distribution in the 15-20, 55-60 and 95-100 h$^{-1}$ Mpc separation bins respectively. The top left, top right, bottom left and bottom right panels display the results for $\Psi_\parallel$, $\Psi_\perp$, $\psi_1$ and $\psi_2$, respectively.}
\end{figure}

Fig.~\ref{fig:CF4-CV_compare} displays the velocity correlation distribution of 100 mock catalogs using these four correlation estimators in three different separation bins, indicating the distribution of the cosmic variance. In the figure, the cosmic variance of $\Psi_\parallel$ and $\psi_2$ follow Gaussian distributions, while the cosmic variance of $\Psi_\perp$ and $\psi_1$ are non-Gaussian, which agrees with correlation result using CF3 catalog \citep{WanRooFel2018}. This suggests that the cosmic variance of the velocity correlation function is a convolution of Gaussian distribution and Wishart distribution. Since $\Psi_\parallel$ and $\Psi_\perp$ can be regarded as combinations of $\psi_1$ and $\psi_2$, their cosmic variance behave similarly that one component ($\Psi_\parallel$ or $\psi_2$) is more Gaussian while the other ($\Psi_\perp$ or $\psi_1$) is more Wishart. The non-Gaussian cosmic variance of $\Psi_\perp$ and $\psi_1$ are present in both the CF3 and CF4 data, indicating that this non-Gaussian cosmic variance may be intrinsic to the velocity correlation functions. Furthermore, this non-Gaussian cosmic variance is unstable across difference separation bins and can impact the cosmological constraints. For $\psi_2$, the situation remains controversial: some studies find that $\psi_2$ is noisier and less stable than $\psi_1$ \citep[e.g.][]{GorDavStr1989, DupCouKub2019}, while others report that $psi_2$ can remain well-behaved \citep[e.g.][]{HelNusFei2017, TurBlaRug2021}. 

\subsection{Observers Effects}\label{sec:result_1}
\begin{figure*}
\centering
\includegraphics[width=5.7cm]{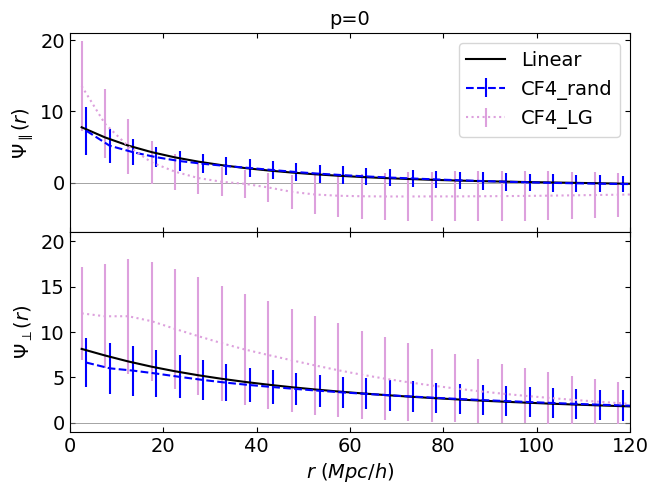}
\includegraphics[width=5.7cm]{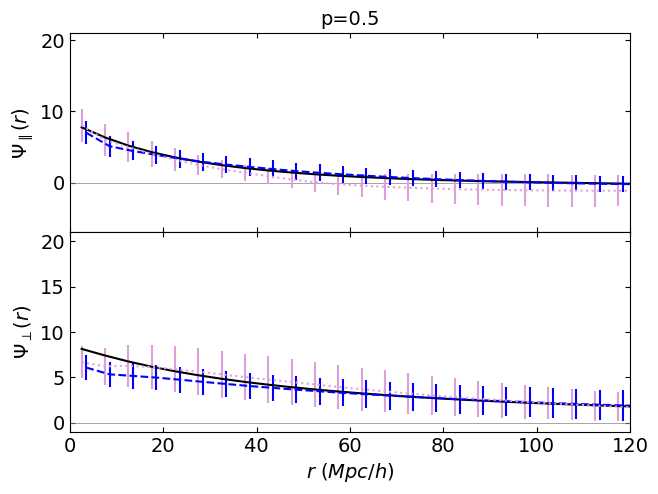}
\includegraphics[width=5.7cm]{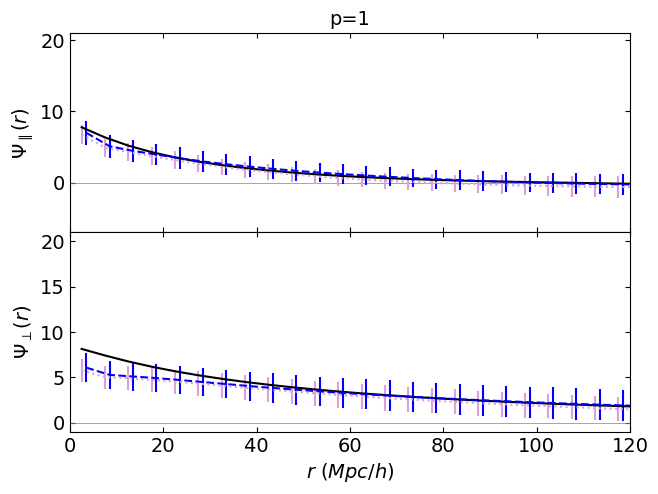}
\caption{\label{fig:correlation_observer} 
The correlation result of the Mask mock catalogs. The left panel indicates correlations with a weight $p=0$, the middle panel shows the result of $p=0.5$, and the right panel represents the result of $p=1$. In each panel, the upper and lower figures illustrates the $\Psi_\parallel$ and $\Psi_\perp$, respectively. The $\Psi_\parallel$ and $\Psi_\perp$ are in unit of $(100$ km s$^{-1})^2$. The blue line shows the result using Rand observers, the plum line represents the result of LG observers, and the black line indicates the linear prediction. The error bars indicate the cosmic variance of the 100 mock catalogs.}
\end{figure*}

The effect of observers on the velocity correlation function was first discussed in \cite{HelNusFei2017} and further improved by implementing the volume-weighted weights introduced in \cite{WanPeeFel2021}. Fig.~\ref{fig:correlation_observer} shows the mask mock correlations for weights $p=$ 0, 0.5 and 1, with error bars representing the cosmic variance. For the CF4 data, the discrepancy caused by different types of observers is not as significant as in the CF3. Particularly for the parallel correlation function, the Rand and LG observers' results agree with each other within the $1\sigma$ cosmic variance even at $p=0$. In contrast, $\Psi_\perp$ is more sensitive to observer types than $\Psi_\parallel$, with the Rand and LG results converging at $p=0.5$. Moreover, $\Psi_\perp$ shows a significantly larger cosmic variance than $\Psi_\parallel$ for both the Rand and LG observers. In addition, the cosmic variance of $\Psi_\perp$ follows a non-Gaussian distribution, as discussed in section~\ref{sec:result_CV}. Given these instabilities of $\Psi_\perp$, we focus our analysis on $\Psi_\parallel$ in this paper. Compared to CF3, the large sample size and sky coverage of CF4 minimize the spatial distribution effect caused by the Virgo Cluster. Consequently, for large peculiar velocity surveys with a fair sample, the observer effect on the velocity correlation functions would be negligible. Given that higher $p$ values assign greater weight to distant galaxies with larger velocity uncertainties which lead to increased statistical uncertainties, we adopt the Rand observer without using the volume-weighted weights ($P=0$) in this paper.

\subsection{Anisotropy Effects}\label{sec:result_2}

\begin{figure}
\centering
\includegraphics[width=8.5cm]{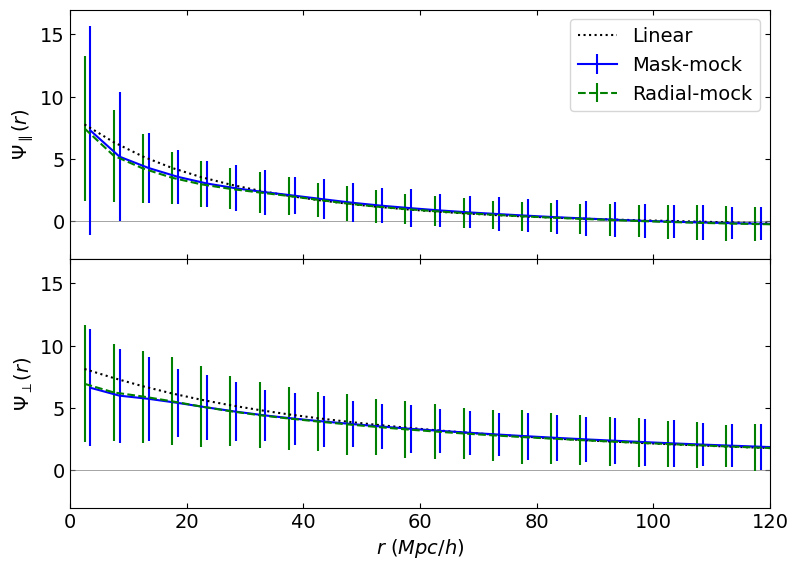}
\caption{\label{fig:correlation_mask} The correlation function of mock catalogs. The upper panel displays the result of $\Psi_\parallel$ in unit of $(100$ km s$^{-1})^2$, while the lower panel represents the result of $\Psi_\perp$. The blue line and the green line denote the results of the CF4 Mask and Radial mocks, respectively. The black dotted line indicates the linear prediction. The error bars show the total uncertainty of the correlation functions.}
\end{figure}

Although CF4 has larger depth, its severe anisotropy in the northern and southern hemispheres may pose new challenges to the estimation of velocity correlation functions. Fig.~\ref{fig:correlation_mask} shows the correlation results using CF4 Mask mocks and CF4 Radial mocks. In the figure, the average correlations of the Mask mock align with the results of the Radial mock. However, the total uncertainty of $\Psi_\parallel$ using Mask mocks is significantly larger than the uncertainty using Radial mocks, particularly in small separation regions. Additionally, the uncertainty of $\Psi_\perp$ is smaller than that of $\Psi_\parallel$ in small separation bins, while the uncertainty of $\Psi_\perp$ becomes larger than that of $\Psi_\parallel$ from the separation bin at 20 $h^{-1}$Mpc.

\begin{figure}
\centering
\includegraphics[width=8.5cm]{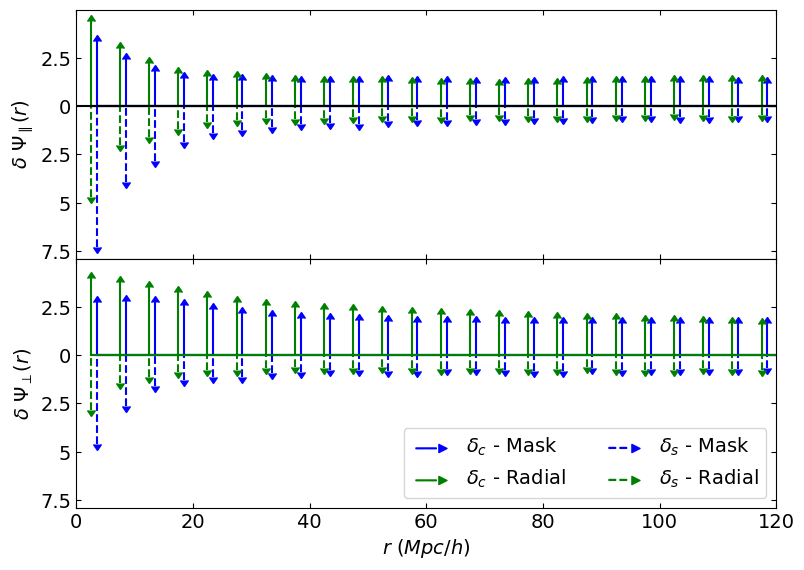}
\caption{\label{fig:correlation_unertainty} The uncertainty of $\Psi_\parallel$ (upper panel) and $\Psi_\perp$ (lower panel) in unit of $(100$ km s$^{-1})^2$. Arrows above $x$ axis represent the cosmic variance ($\delta_c$) of $\Psi_\parallel$, while arrows below the axis indicate the statistical uncertainty ($\delta_s$) of $\Psi_\parallel$. The blue arrows display the result using the Mask mocks and the green arrows show the result of the Radial mocks.}
\end{figure}

Fig.~\ref{fig:correlation_unertainty} resolves the total uncertainty ($\delta_t$) into cosmic variance ($\delta_c$) and statistical uncertainty ($\delta_s$) using the DS peculiar velocity estimator. The figure shows the difference between the correlation results of Mask and Radial mocks is primarily due to the statistical uncertainty. Compared to the CF3 velocity correlation function \citep{WanRooFel2018, WanPeeFel2021}, the CF4 velocity correlation function exhibits a significantly increased statistical uncertainty, which was previously smaller than the cosmic variance in CF3. The large statistical uncertainty of the CF4 correlation function can be attributed to two main factors: 1) the unbalanced sample distribution in CF4, which leads to larger statistical uncertainties; 2) the substantial velocity uncertainty associated with the large depth of CF4, which significantly increases the statistical uncertainty. 

In CF3 correlation studies \citep[e.g.][]{WanRooFel2018, WanPeeFel2021}, the anisotropy showed no significant effect on the average value and uncertainties of correlation functions. The severe effect of anisotropy on CF4 correlations might arise from the fact that the anisotropy in CF4 data is not only angular but also radial, which the depth of northern sky (z$\sim$0.1) is two times of the depth of southern sky (z$\sim$0.05). Furthermore, the similar cosmic variance between the Mask and Radial mocks reinforces the conclusion that the effect of different types of observers becomes negligible for large peculiar velocity surveys.

\subsection{Weights} 
\label{sec:weight}

In section~\ref{sec:result_1}, we discuss the effect of correlation weights on the bias of cosmic variance caused by the observer effect. In this section, we implement three different weighting schemes to explore the effect of correlation weights on statistical uncertainties: the FKP weight \citep{FKP1994}, revised FKP weight \citep{Howlett2019} the modified kernel density weight.

For the FKP weighting scheme, the weight $w$ in eq.~\ref{eq:psipar} and ~\ref{eq:psiperp} can be expressed by $w=W_{FKP}^i\cdot W_{FKP}^j$, where the superscripts $i$ and $j$ indicate the two groups in a group pair. The $W_{FKP}$ represents the FKP weight derived from eq.~\ref{eq:FKP}, where $\overline{n}(r)$ is the average number density of groups at radial distance $r$, and P(k) denotes the density power spectrum at certain $k$. In this paper, we set $P(k)=1600$ $h^{-3}$ Mpc$^3$ following \citet{QinHowSta2019}.

For the revised FKP weighting scheme, the weight is given by $w=W_{FKP_v}^i\cdot W_{FKP_v}^j$ as shown in eq.~\ref{eq:FKPv}. In this expression, $\sigma_v$ denotes the velocity uncertainty and $P_v(k)=10^9$ $h^{-3}$ Mpc$^3$ km$^2$ s$^{-2}$ following \citet{TurBlaRug2023}.

\begin{align}
W_{FKP}(r) &= \frac{1}{1+\overline{n}(r)P(k)} \label{eq:FKP}\\
W_{FKP}(r) &= \frac{1}{\sigma_v+\overline{n}(r)P_v(k)} \label{eq:FKPv}\\
W_\rho &= \sqrt{\frac{cz}{cz^{L}_{CF3}}}\frac{\rho}{\overline{\rho}} \label{eq:kernel_density}
\end{align}

For the modified kernel density weighting scheme, the correlation weight is expressed as $w=W_\rho^i\cdot W_\rho^j$, where $W_\rho$ indicates the modified kernel density weight given by eq.~\ref{eq:kernel_density}. In the equation, $\rho$ represents the kernel density at the position of each groups, $\overline{\rho}$ is the average kernel density across over groups, $cz$ is the redshift of each group in km s$^{-1}$, and $cz^{L}_{CF3}$ is the redshift limit of CF3 dataset, which is approximately 15,000 km s$^{-1}$. Due to the imbalanced group distribution between regions inside and outside the redshift of 15,000 km s$^{-1}$, we implement $\sqrt{\frac{cz}{cz^{L}_{CF3}}}$ to assign more weight to outside groups, thereby balancing the kernel density weight.

\begin{figure}
\centering
\includegraphics[width=8.5cm]{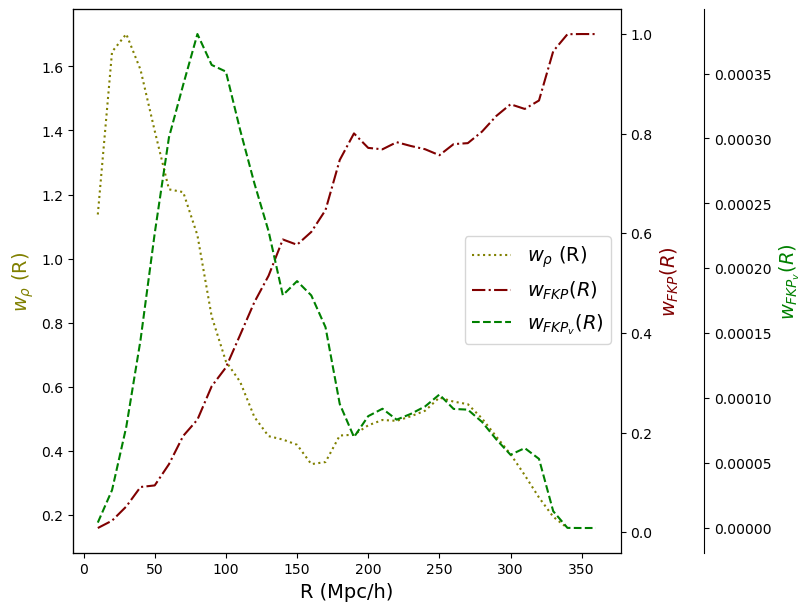}
\caption{\label{fig:window_function} Weight function of redshift from the CF4 group catalog. The maroon dash-dotted line indicates the FKP weight with magnitude represented by the marron y-axis on the right-hand side. The green dashed line represents the revised FKP weight with magnitude represented by the green y-axis on the right-hand side. The olive dotted line denotes the modified kernel density weight with magnitude displayed by the y-axis on the left-hand side.}
\end{figure}

Fig.~\ref{fig:window_function} shows the FKP weight, revised FKP weight and modified kernel density weight as functions of redshift. In the figure, the FKP weight increases with the redshift of groups, while the revised FKP weight and the modified kernel density weight decreases with the redshift after 100 $h^{-3}$ Mpc$^3$. Therefore, the FKP weighting scheme assigns more weight to distant groups, while the revised FKP weight and the modified kernel density weighting scheme gives more weight to nearby groups.

\begin{figure}
\centering
\includegraphics[width=8.5cm]{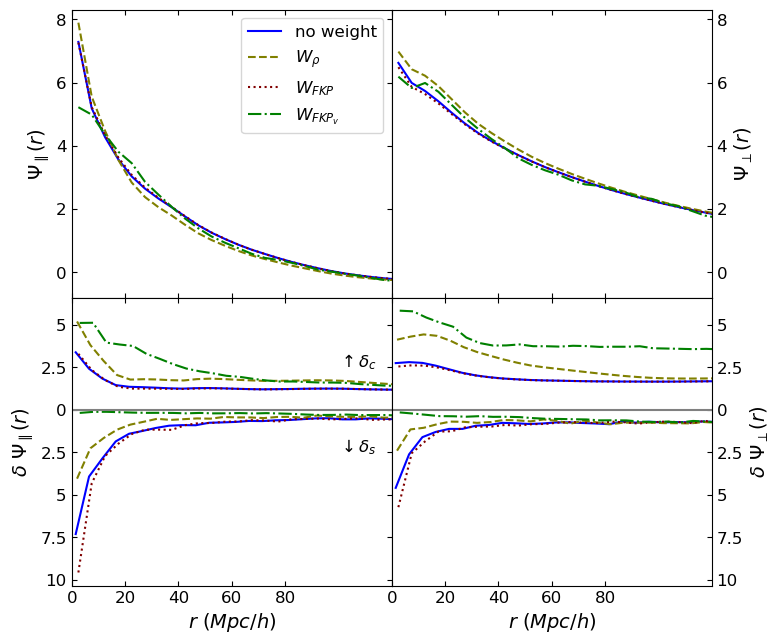}
\caption{\label{fig:weight} Mean and uncertainties of the weighted $\Psi_\parallel$ (upper panel) and $\Psi_\perp$ (lower panel) of the CF4 Mask mock catalogs, in unit of $(100$ km s$^{-1})^2$. The top panels show the mean correlation functions, while the bottom panels indicate the associated uncertainties. The blue solid line illustrates the correlation without weight, the olive dashed line denotes the correlation with the modified kernel density wight, the marron dotted line indicates the result using the FKP weight, the green dash-dotted line indicates the result using the revised FKP weight. In the bottom panels, lines above $y=0$ represent the cosmic variance ($\delta_c$), while lines below indicate the statistical uncertainty ($\delta_s$).}
\end{figure}

Fig.~\ref{fig:weight} shows the mean and uncertainty of correlation functions using different weighting schemes. In the figure, the statistical uncertainties of both $\Psi_\parallel$ and $\Psi_\perp$ decrease significantly with the revised FKP weight and the modified kernel density weight, while the statistical uncertainties using the FKP weighting scheme show no significant difference from the correlations without weighting. The statistical uncertainty of velocity correlation function is caused by the velocity uncertainties. Since the velocity uncertainty increases with distance, assigning greater weight to nearby groups, where these velocity uncertainties are smaller, is essential for reducing statistical uncertainties.

However, assigning larger weights to nearby groups would lead to a larger cosmic variance. As discussed in Section~\ref{sec:result_CV}, the cosmic variance depends on the survey depth, with deeper surveys generally exhibiting smaller cosmic variance. Therefore, weighting schemes that up-weight nearby groups (such as the revised FKP weight and the modified kernel density weight) might reduce the effective depth region of CF4 dataset, thereby increasing the cosmic variance. In this sense, the revised FKP weight may not be an optimal weighting scheme for the CF4 velocity survey. However, future surveys are expected to be both deeper and denser, such as the Dark Energy Spectroscopic Instrument (DESI) peculiar velocity survey \citep{SauHowDou2023, BauAmsAro2025, DouBenKim2025, RosHowLuc2025}, which would reduce the impact of cosmic variance. In that regime, the revised FKP weight would remain promising, as its ability to suppress statistical uncertainties may become more advantageous.

In contrast, the FKP weighting scheme assigns more weight on distant groups, whose peculiar velocity uncertainties are extremely large. With more weigh on the distant groups, the large velocity uncertainties dominate the statistical uncertainty of the velocity correlation function. Therefore, the statistical uncertainty using the FKP weights show no significant difference from the correlations without weighting. In addition, the standard FKP weight was designed for density tracers and does not explicitly include velocity information, which limits its effectiveness for the velocity correlation measurement.

\begin{figure}
\centering
\includegraphics[width=8.5cm]{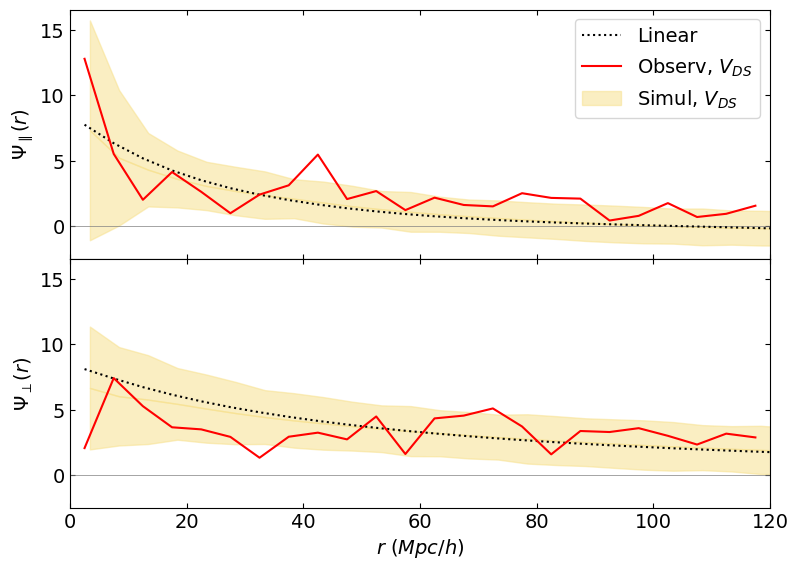}
\caption{\label{fig:correlation_observation} The $\Psi_\parallel$ (upper panel) and $\Psi_\perp$ (lower panel) of the CF4-group observation catalog in unit of $(100$ km s$^{-1})^2$. The red solid represents the observational correlation function with the DS peculiar velocity estimator, the blue dash-dotted line shows the observational result with the Hubble peculiar velocity estimator, and the green dashed line indicates the result using the modified estimator. The black dotted line denotes the correlation result predicted by linear theory. The contours display the correlation function and total uncertainties of the corresponding mock catalogs using the DS peculiar velocity estimator.}
\end{figure}

Given the impact of weighting on the CF4 correlation uncertainties, we do not apply any weights in the following analysis of the CF4 velocity correlation function and cosmological parameter constraints. Fig.~\ref{fig:correlation_observation} displays the correlation results from the CF4-group observation catalog using DS estimators without weight.

\section{Constraints}
\label{sec:constraints}

With peculiar velocity measurements extended to $z\sim 0.1$, the CF4 survey provides a broader view for peculiar velocity studies. In addition, the deeper and larger coverage of the CF4 dataset offers reduced and more stable cosmic variance, while the DS peculiar velocity estimator improves the statistical uncertainty. Under these conditions, we implement the Markov Chain Monte Carlo (MCMC) method to constrain $f\sigma_8$.

In the MCMC analysis, the parameter set $\lbrace f\sigma_8 \rbrace$ is treated as variables in the linear theory prediction ($\Psi_\parallel^L$). The likelihood function is derived by fitting the observed CF4 parallel correlation function ($\Psi_\parallel^O$) with the linear theory prediction ($\Psi_\parallel^L$):
\begin{equation}\label{eq:likelihood}
    \mathcal{L} = -\frac{1}{2}\sum_{i,j}\left[ \psi^O_{\parallel}(r_i) - \psi^L_{\parallel}(r_i)\right] (\alpha C^{-1}_{ij})\left[ \psi^O_{\parallel}(r_j) - \psi^L_{\parallel}(r_j)\right],
\end{equation}
where $\alpha=\frac{n-p-2}{n-1}$ is the Hartlap factor \citep{HarSimSch2007} with $n=100$ mocks and $p=17$ separation bins covering $20-100$ $h^{-1}$Mpc; $C_{ij}$ is the covariance matrix derived by using both the total correlation uncertainty ($\delta_t$) and the statistical correlation uncertainty ($\delta_s$): 
\begin{equation}\label{eq:ET_covariance_matrix}
C_{ij}^{\delta_t} = \frac{1}{N^t_{mock}} \sum^{N^t_{mock}}_{n=1} \left( \Psi^i_{\parallel, n} - \overline{\Psi}^i_{\parallel, t} \right) \left( \Psi ^j_{\parallel, n} - \overline{\Psi}^j_{\parallel, t}\right)^T,
\end{equation}
\begin{equation}\label{eq:ST_covariance_matrix}
C_{ij}^{\delta_s} = \frac{1}{N^s_{mock}} \sum^{N^s_{mock}}_{n=1} \left( \Psi^i_{\parallel, m} - \overline{\Psi}^i_{\parallel, s} \right) \left( \Psi ^j_{\parallel, m} - \overline{\Psi}^j_{\parallel, s}\right)^T.
\end{equation}

In Eq.~\ref{eq:ET_covariance_matrix}, $N^t_{mock}$ is the number of perturbed mock catalogs, obtained by randomly perturbing each of the 100 mock catalogs once according to the DS velocity uncertainties; $\psi^i_{\parallel, n}$ represents the correlation value in the $i^{th}$ separation bin of the $n^{th}$ perturbed mock catalog; $\overline{\psi}^i_{\parallel, t}$ indicates the average correlation value across all $N^t_{mock}$ perturbed mock catalogs in the $i^{th}$ separation bin.  In Eq.~\ref{eq:ST_covariance_matrix}, $N^s_{mock}$ is the number of perturbed catalogs of selected simulation mock whose value is close to the average correlation function; $\psi^i_{\parallel, m}$ represents the correlation value in the $i^{th}$ separation bin of the $m^{th}$ perturbation catalog; $\overline{\psi}^i_{\parallel, s}$ indicates the average correlation of the perturbed catalog of the selected simulation mock.

Since the covariance matrix is calculated using mock catalogs from OuterRim Simulation, it is fixed to the fiducial cosmology of OuterRim (WMAP7). While fixing the covariance matrix to a fiducial cosmology can, in principle, affect the constrained parameter uncertainties \citep[e.g.][]{BloCorRas2021}, this effect can be negligible when the fiducial cosmology is close to the the best-fit values. Additionally, given the relatively large uncertainties in our parameter constraints, the impact of using a fixed covariance matrix is negligible in our analysis.

In this paper, we focus solely on constraining $f\sigma_8$, since including additional parameters such as the Hubble constant $H_0$, the Baryon density $\Omega_b$ or the Primordial scalar spectral index $n_s$ would cause severe degeneracies, resulting in unreliable constraints. In addition, constraining $\Omega_m$ and $\sigma_8$ independently reveals a strong degeneracy, which may be due to the limit of using CF4 data to constrain $\Omega_m$ with its restricted redshift range.

We set $\Omega_m=0.2648$, $\Omega_b=0.0448$ and $n_s=0.963$, consistent with the OuterRim Simulation (Table~\ref{T_Mill}), and $H_0=74.6$ (km s$^{-1}$ Mpc$^{-1}$) from CF4 \citep{TulKouCou2023}. We run the MCMC analysis using the emcee \citep{ForHogLan2013} with 100 walkers and 100,000 iterations for both the statistical uncertainties ($\delta_s$) case and total uncertainties ($\delta_t$) case. The walkers are initialized in a small Gaussian ball around the starting point $f\sigma_8=0.4$, within the uniform prior ranges $f\sigma_8 \in (0.05,0.99)$. We adopt a mixture of moves comprising the StetchMove ($70\%$) and the DEMove ($30\%$). For the final parameter constraints, we retain the last 10,000 iterations ($L_i=10,000$, retained iteration length) with all the walkers in each case.

\begin{figure}
\centering
\includegraphics[width=8.5cm]{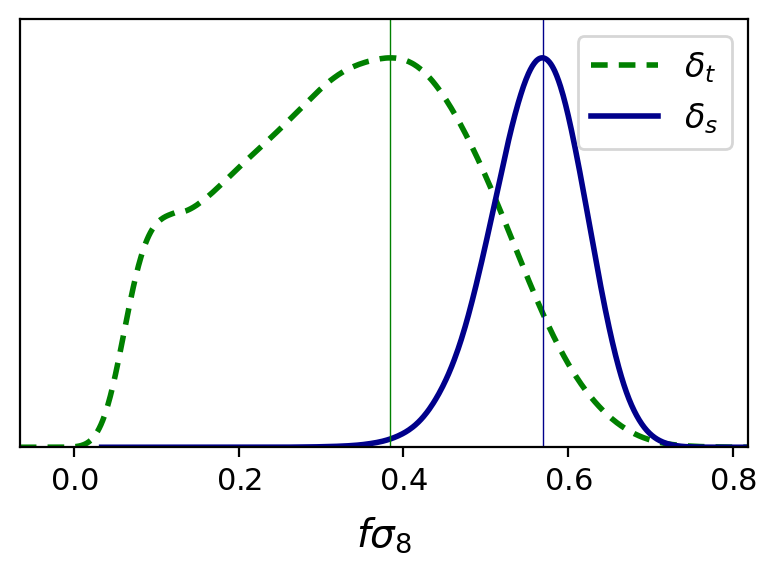}
\caption{\label{fig:fs8} The constraints on $f\sigma_8$ derived from $\Psi_\parallel$ of CF4-group observational data.}
\end{figure}

Fig.~\ref{fig:fs8} shows the constraints on $f\sigma_8$ derived from the CF4-group observational parallel correlation function ($\Psi_\parallel$) over separations ranging from 20 to 100 $h^{-1}$Mpc. Due to the large uncertainties at small separations, we exclude the first three separation bins and begin the cosmological parameter constraints from the fourth bin (20 $h^{-1}$Mpc). On the large-scale end, we explore truncation boundaries from 80 to 120 $h^{-1}$Mpc, and find that the results converge starting at 100 $h^{-1}$Mpc. Therefore, we adopt a separation range of 20 to 100 $h^{-1}$Mpc for our analysis.

Using the MCMC method, we obtain $f\sigma_8^T=0.384^{+0.116}_{-0.194}$ from the CF4 parallel velocity correlation function with total correlation uncertainties ($\delta_t$), while the results using statistical uncertainties ($\delta_s$) are $f\sigma_8^S=0.569^{+0.054}_{-0.06}$ (Table~\ref{T_param}). Compared to Planck's results \citep{PlanckAgh2019} of $\Omega_m^P=0.3153\pm 0.0073$ and $\sigma_8^P=0.8111\pm 0.006$, our constraints for $f\sigma_8$ using total uncertainties are consistent with Planck's results within $1\sigma$. However, the uncertainty of the $f\sigma_8^T$ is relatively large, which may arise because the covariance matrix is affect by the large uncertainties in the velocity correlation function. For constraints using statistical uncertainties, the value of $f\sigma_8$ is noticeably higher and the constraining becomes tighter. However, using only the statistical uncertainty would effectively limits the analysis to the local universe. If the local value of $\Omega_m$ is lower than the value inferred from WMAP7, it could introduce a bias to our local $f\sigma_8$ constraining result.

Table~\ref{T_param} summarizes the MCMC constraint statistics. The integrated autocorrelation time, $\tau$, quantifies the correlation length of the MCMC chains and provides an estimate of the number of iterations required to obtain approximately independent posterior samples. It is therefore widely used as a diagnostic of convergence. In practice, the retained iteration length, $L_i$, is required to be much larger than $\tau$ (typically $\geq50\tau$) to ensure convergence. In our analysis, we retain $L_i=10,000$ iterations, which is substantially larger than the $\tau$ values reported in Table~\ref{T_param}. We therefore conclude that the MCMC constraints have converged for both the statistical uncertainty and total uncertainty cases. 

However, the reduced chi-square values are larger than unity in both cases. For the statistical uncertainty case, we obtain $\chi^2_\nu=2.0$, which may reflect a combination of effects: 1) statistical uncertainties derived solely from velocity measurement errors may not capture additional sources of statistical uncertainty; 2) the measured CF4 $\Psi_\parallel$ exhibits noticeable bin-to-bin fluctuations (Fig.~\ref{fig:correlation_observation}), which can influence the fitting;  3) the non-linearity effect may impact the results when fitting with linear theory predictions. For the total uncertainty case, we obtain $\chi^2_\nu=2.2$, which is comparatively high. In addition to the effect listed above, this may also be related to the non-Gaussian distributions of cosmic variance. Although the cosmic variance of $\Psi_\parallel$ is expected to be more nearly Gaussian than that of $\psi_1$, the deviations from Gaussianity may still remain and contribute to the elevated $\chi^2_\nu$. We expect that with larger datasets, such as DESI, the effects of the bin-to-bin fluctuations and the correlation uncertainty will be reduced, leading to improved parameter constraints and a lower reduced chi-square velue.

\begin{table}
\caption{MCMC summary statistics. The result with label $\delta_s$ and $\delta_t$ correspond to constraints using the statistical uncertainty and total uncertainty covariance matrices, respectively. $\chi^2_\nu$ represents the reduced chi-square, and $\tau$ is the integrated autocorrelation time.}
\centering
{
\begin{tabular}{lccc}
\hline
            & $f\sigma_8$ & $\chi^2_\nu$ & $\tau$ \\
\hline
$\delta_s$  & $0.569^{+0.054}_{-0.06}$ & 2.0 & 20.2 \\
\\
$\delta_t$  & $0.384^{+0.116}_{-0.194}$ & 2.2 & 27.2 \\
\hline
\end{tabular}%
}
\label{T_param}
\end{table}

Compared to $f\sigma_8^S$, the posterior distribution of $f\sigma_8^T$ is non-Gaussian. This arises from the structure of the total uncertainty, which includes contributions from cosmic variance. Although the cosmic variance of $\psi_\parallel$ is more Gaussian than that of the Gorski correlation, its distribution still indicates a convolution of Gaussian and Wishart distributions. This non-Gaussian component leads to the non-Gaussian posterior distribution for $f\sigma_8^T$.

We compare our result with previous growth rate measurements derived from velocity correlation functions. \citet{DupCouKub2019} constrained growth rate using CF3 data with a fixed power spectrum using Planck 2015 cosmology, treating $f\sigma_8$ as the sole free parameter. They obtained a local growth rate (using statistical uncertainty) of $f\sigma_8^S=0.43\pm0.03$ and a cosmic growth rate (using cosmic variance) of $f\sigma_8^C=0.43\pm0.11$. \citet{TonAppPar2024} reported a local growth rate of $f\sigma_8^S=0.361^{+0.02}_{-0.027}$ and cosmic growth rate of $f\sigma_8^C=0.361^{+0.013}_{-0.015}$ using simulation data with a survey depth of 500 $h^{-1}$Mpc and a Gaussian velocity uncertainty of 3000 km s$^{-1}$. Additionally, \citet{CouDupGui2023} constrained a local growth rate of $f\sigma_8^S=0.32\pm0.07$ using CF4 group data.

Our measured growth rate ($f\sigma_8^T=0.384^{+0.116}_{-0.194}$) is consistent with the result of \citet{CouDupGui2023}, despite differences in the correlation estimators used, as both analyses are based on the same CF4 dataset. The uncertainty of our measured growth rate is significantly larger than that of \citet{CouDupGui2023} and that obtained using the CF3 data. This increase arises from the inclusion of cosmic variance as well as the large statistical uncertainties driven by the substantial velocity uncertainties of the newly added distant galaxies in CF4. Furthermore, \citet{TonAppPar2024} adopted a velocity uncertainty of 3000 km s$^{-1}$, which is significantly lower than the velocity uncertainties of CF4 (Figure~\ref{fig:CF4-v_error}).

\section{Conclusion}
\label{sec:conclusion}

In previous studies, the measurement and estimation of the velocity correlation function was limited by the sample size and data precision of peculiar velocity surveys. For example, the velocity correlation function varied with different types of observes \citep{HelNusFei2017}. In our previous work \citep{WanPeeFel2021}, we introduced a position-weighted method to mitigate the bias caused by the type of observers. The compilation of the new peculiar velocity survey, CF4, was expected to address some of the shortcomings of the studies of velocity correlation functions, while also presenting new challenges.

In this paper, we analyze the anisotropy of the CF4 data, which exhibits an unbalanced distribution of galaxies between the northern and southern hemispheres. The CF4 survey contains more galaxies in the northern hemisphere, where all the distant galaxy additions from redshift 0.05 to 0.1 are located. Given the severe anisotropy of the CF4 data, the mock catalogs are generated using a mask that characterizes the anisotropy distribution in the CF4.

We apply the parallel ($\Psi_\parallel$) and perpendicular ($\Psi_\perp$) correlation functions introduced by \cite{WanPeeFel2021} to the CF4 data and mock catalogs, and test the cosmic variance of four velocity correlation function estimators using the CF4 data. The estimators $\Psi_\parallel$ and $\psi_2$ exhibit Gaussian-distributed cosmic variance, whereas the cosmic variance of $\Psi_\perp$ and $\psi_1$ fellow non-Gaussian distributions. Considering the performances of these four estimators, $\Psi_\parallel$ remains the optimal velocity correlation estimator for the CF4 data.

Our results indicate that the effect of observer on the cosmic variance of velocity correlation functions becomes negligible at the extensive survey depth of CF4. However, the severe anisotropy caused by the uneven sample distribution between the northern and southern hemispheres, along with the large velocity uncertainty of distant groups, leads to increased statistical uncertainty.

To reduce the uncertainty in the CF4 velocity correlations, we test the impact of weighting schemes on statistical uncertainties. Both the revised FKP weight and the modified kernel density weight reduce the statistical uncertainty, but increase the cosmic variance. For the CF4 dataset, these two schemes tend to up-weight nearby groups, reducing the effective survey depth of CF4 and thereby enhancing the impact of cosmic variance. However, this limitation is expected to be marginalized in future peculiar velocity surveys that are deeper and denser, since the effect of cosmic variance decreases with survey depth.

To constrain cosmological parameters under large velocity correlation uncertainties, we implement the MCMC analysis. The MCMC provides cosmological constraints of $f\sigma_8^T=0.384^{+0.116}_{-0.194}$ using total correlation uncertainties, and local cosmological constraints of $f\sigma_8^S=0.569^{+0.054}_{-0.06}$ using statistical correlation uncertainties. The constraints using total uncertainties are relatively loose, primarily due to the large uncertainties in the velocity correlation functions caused by the substantial velocity uncertainties. For the local constraints, we obtain a relatively lower value with smaller uncertainty.

With deeper and denser peculiar velocity surveys on the horizon, such as the DESI Peculiar Velocity survey, peculiar velocity studies are expected to provide increasingly improved cosmological constraints.

\section*{Acknowledgements}

This work is supported by the National Key R\&D Program of China (2023YFA1607800, 2023YFA1607804), 111 project No. B20019, and Shanghai Natural Science Foundation, grant No.19ZR1466800. We acknowledge the science research grants from the China Manned Space Project with No.CMS-CSST-2021-A02. HAF was partially supported by US National Science Foundation grant AST-1907404. RW was partially supported by the US National Science Foundation grant AST-1907365. 
The computations in this paper were run on the Gravity Supercomputer at Shanghai Jiao Tong University.

\section*{Data Availability Statement}

The data underlying this article will be shared on reasonable request to the corresponding author.

\bibliographystyle{mnras}
\bibliography{Yuyu}

\bsp	
\label{lastpage}
\end{document}